\documentclass[pra,aps,twocolumn,superscriptaddress,showpacs,showkeys, longbibliography]{revtex4-1}
\usepackage{calc,amsmath,amsthm,amssymb,graphicx,subfigure,xcolor,comment}
\usepackage{mathdots}
\usepackage[T1]{fontenc}
\usepackage[unicode=true]{hyperref}
\usepackage{booktabs}
\usepackage{threeparttable}

\hypersetup{
     colorlinks=true,       		% false: boxed links; true: colored links
     linkcolor=blue,          	% color of internal links
     citecolor=red,            % color of links to bibliography
     urlcolor=magenta,           	% color of external links
 }
\newtheorem*{theorem*}{Theorem}

\newtheorem*{corollary*}{Corollary}

\newtheorem*{lemma*}{Lemma}

\newtheorem*{proposition*}{Proposition}
\theoremstyle{definition}

\newtheorem*{definition*}{Definition}
\theoremstyle{remark}

\newtheorem*{remark*}{Remark}
\newcommand{\ket}[1]{|#1\rangle}
\newcommand{\bra}[1]{\langle#1|}

\begin{document}
\renewcommand{\figurename}{Fig.}

\title{Two Quantum Entangled Cheshire Cats}

\author{Jie Zhou}
\affiliation{Theoretical Physics Division, Chern Institute of Mathematics, Nankai University, Tianjin 300071, People's Republic of China}
\affiliation{Centre for Quantum Technologies, National University of Singapore, 3 Science Drive 2, Singapore 117543}

\author{Leong-Chuan Kwek}
\email{kwekleongchuan@nus.edu.sg}
\affiliation{Centre for Quantum Technologies, National University of Singapore, 3 Science Drive 2, Singapore 117543}
\affiliation{MajuLab, CNRS-UNS-NUS-NTU International Joint Research Unit, UMI 3654, Singapore}
\affiliation{National Institute of Education, Nanyang Technological University, 1 Nanyang Walk, Singapore 637616}
\affiliation{School of Electrical and Electronic Engineering Block S2.1, 50 Nanyang Avenue, Singapore 639798}

\author{Jing-Ling Chen}
\email{chenjl@nankai.edu.cn}
\affiliation{Theoretical Physics Division, Chern Institute of Mathematics, Nankai University, Tianjin 300071, People's Republic of China}

\date{\today}
\begin{abstract}
Quantum entanglement serves as an important resource for quantum processing.
In the original thought experiment of the Quantum Cheshire Cat, the physical properties of the cat (state) can be decoupled from its quantum entities. How do quantum entanglement and weak values affect such thought experiment? Here, we conceive a new thought experiment that exploits quantum entanglement with weak (value) measurement. Specifically, we ask: can two entangled particles exchange physical and quantum properties? We answer the question in the affirmative.
\end{abstract}



\maketitle

\section{Introduction}

There are ostensibly many counter-intuitive phenomena in the quantum world: the Quantum Cheshire Cat thought experiment is one of such  examples ~\cite{aharanov2005quantum,bell1964einstein,brunner2014bell,kochen1975problem,budroni2021quantum,leggett1985quantum,Aharonov2013QuantumCats}. It describes a scenario in which the properties, such as the polarization of a photon, can exist away from the photon itself. This is counter-intuitive to our macro-intuition since in real life, the properties of an object are attached to the object itself. Thus, we cannot discuss the nature of the object in isolation with itself, since such possibility does not make sense. Indeed, the name of Quantum Cheshire Cat comes from the famous novel <<Alice in Wonderland>>~\cite{carrollalice}.
In this novel, there is a magic cat, ```Well! I've often seen a cat without a grin', thought Alice, `but a grin without a cat! It's the most curious thing I ever saw in my life!''' In 2013, Yakir Aharonov {\it et. al.} first proposed the Quantum Cheshire Cat phenomenon for a quantum setting. They use this story to visualize the separation of attributes from matter, i.e., the grin can leave the Quantum Cheshire Cat~\cite{Aharonov2013QuantumCats}.

Quantum measurement has always been one of the important, yet subtle, problems in all quantum theories~\cite{braginsky1995quantum,wiseman2009quantum}. Although quantum physics has been proven correct, at the philosophical level, there has always been a debate about the process of measurement.
The implementation of the thought experiment on the Quantum Cheshire Cat relies primarily on weak values~\cite{Aharonov2015WeakNonlocality,aharonov1990properties,aharonov2005new,jozsa2007complex,hosten2008observation,lundeen2011direct}. %The weak measurement can ensure that the measurement does not disturb the initial state much.
Here, we introduce the concept of weak values~\cite{Aharonov2015WeakNonlocality}.
The initial state of a system is denoted as $\ket{\Psi_0}$, while the final state of the system is denoted as $\ket{\Psi_f}$.
In quantum mechanics, the initial and final states of the system are known as the pre- and post-selected quantum states, the weak value of any observable $O$ is given by
\begin{align}\label{eq:weak}
\langle O\rangle_w=\frac{\bra{\Psi_f}O\ket{\Psi_0}}{\bra{\Psi_f}\Psi_0\rangle}.
\end{align}
%where $\ket{\Psi_0}$ and $\ket{\Psi_f}$ are the pre-selected state and post-selected state respectively.
The weak value of an observable gives the strength of the coupling between the observable operator and the quantum evolution system.
There are some fascinating aspects for weak values.
From Eq.~(\ref{eq:weak}), it is seen that generally the weak value of an observable is a complex number~\cite{jozsa2007complex}. If the pre-selected state or the post-selected state is the eigenstate of the observable $O$, then, the weak value will be an eigenvalue corresponding to this eigenstate, this is commonly referred to as a strong measurement. When the pre- and post-selected states are two nearly orthogonal states, the weak value can assume large values and exceed the range of the eigenvalue spectrum~\cite{Aharonov2015WeakNonlocality,duck1989sense}. Hence, it can be used effectively in signal amplification~\cite{dixon2009ultrasensitive,nakamura2012evaluation}. Weak measurements have been applied to the spin Hall effect~\cite{hosten2008observation}, Hardy's paradox~\cite{aharonov2002revisiting}, the three-box paradox~\cite{albert1985curious}, measurement of wave-function for a single photon~\cite{lundeen2011direct}, quantum state tomography~\cite{hofmann2010complete,wu2013state}, quantum precision thermometry~\cite{pati2020quantum} and so on.

The physical object and properties can be traced through an association of the positions of observables with their corresponding weak values~\cite{danan2013asking}.
The weak value $\langle O\rangle_w=0$ means that the property represented by the observable is not present. While, the weak value $\langle O\rangle_w=1$ indicates that the corresponding property by the observable is observed. Thus, via weak values, we can detect the location of objects and their properties, in order to determine whether there is a Quantum Cheshire Cat phenomenon.
On the contrary, Quantum Cheshire Cat provides a new way to understanding the ontology of measurement observables.

Quantum Cheshire Cat has attracted a lot of attention since it was proposed~\cite{ibnouhsein2012twin,guryanova2012complete,di2012hunting,matzkin2013three,denkmayr2014observation,bancal2014isolate,correa2015quantum,atherton2015observation,ashby2016observation,duprey2018quantum,das2019teleporting,das2020can,liu2020experimental,das2021delayed,aharonov2021dynamical,hance2022dynamical}.
In 2020, Debmalya Das and Arun Kumar Pati proposed a scheme which has two Quantum Cheshire Cats~\cite{das2020can}. Two photons can be separated with their polarizations and capture others polarization. This phenomenon was soon verified experimentally~\cite{liu2020experimental}.
Apart from the photon and its polarization, the effect of Quantum Cheshire Cat has been shown in experiments with the neutron's magnetic moment~\cite{denkmayr2014observation}.
%Both of them have been observed experimentally.
More generally,  with the appropriate choice of the pre- and post-selected quantum states, we observe the effect of Quantum Cheshire Cat in principle for any quantum system of objects and its properties.
Interestingly, there are some intrinsic connections between the Quantum Cheshire Cat and other quantum paradox, for example, the three-box paradox can be analyzed through the effect of Quantum Cheshire Cat~\cite{albert1985curious,matzkin2013three}.
Otherwise, based on the technology of Quantum Cheshire Cat, some interesting new results could emerge. For example, Pratyusha Chowdhury {\it et. al.} found that wave and particle attributes of a quantum entity can be completely separated, they successfully dismantle the wave–particle duality for a quantum entity~\cite{chowdhury2021wave}.

%Quantum Cheshire Cat has opened up a new window for understanding of quantum systems

In this paper, we put forward a new two Quantum Cheshire Cats scheme. Two initially entangled Quantum Cheshire Cats are found to be separated from their ``grin". Quantum entanglement is an important property in quantum theory~\cite{horodecki2009quantum}. Entangled Quantum Cheshire Cats can further probe the nuance of quantum theory.

The paper is organized as follows. In Sec.~\ref{1cc}, we  briefly review the original scheme of the Quantum Cheshire Cat. In Sec.~\ref{2ecc}, we present our protocol with two entangled Quantum Cheshire Cats and show how they can lose their ``grins". In Sec.~\ref{setup}, we describe how to implement the thought experiment of two entangled Quantum Cheshire Cats. In Sec.~\ref{general}, we extend our scheme to a more general entangled Quantum Cheshire Cats scenario. In Sec.~\ref{necc}, we generalize to an $n$ entangled Quantum Cheshire Cats scenario.
Finally, we conclude and discuss our paper.

\section{The original Quantum Cheshire Cat}\label{1cc}

In this section, we review the original Quantum Cheshire Cat~\cite{Aharonov2013QuantumCats}. To separate the polarization property from a photon,  the pre-selected state should be selected as
\begin{align}
\ket{\psi_0}=\frac{1}{\sqrt{2}}(i\ket{L}+\ket{R})\ket{H},
\end{align}
where $\ket{L}$ and $\ket{R}$ denote two orthogonal states representing the two possible paths taken by the photon, $\ket{H}$ denotes the horizontal polarization state of the photon.
It's obviously that the pre-selected state $\ket{\psi_0}$ is a separable state of two degrees of freedom: location and polarization.

The post-selected state should be selected as
\begin{align}
\ket{\psi_f}=\frac{1}{\sqrt{2}}(\ket{L}\ket{H}+\ket{R}\ket{V}),
\end{align}
where $\ket{V}$ is the vertical polarization state of the photon, which is orthogonal to the vertical polarization state $\ket{H}$. Obviously, this is a entangled state of location and polarization.

The path (``cat'') observables, revealing the spatial positions of the particle (cat), read
\begin{align}
\Pi_{\mu}=\ket{\mu}\bra{\mu},
\end{align}
where $\mu\in\{L, R\}$, $L$ and $R$ denoting the two possible paths (left and right arms) that a particle (Cheshire Cat) can go through.

The circular polarization observable is given by the Pauli
operator
\begin{align}
\sigma_z=\ket{\uparrow}\bra{\uparrow}-\ket{\downarrow}\bra{\downarrow},
\end{align}
%\begin{align}
%\sigma_z=\ket{+}\bra{+}-\ket{-}\bra{-},
%\end{align}
where the circular polarization basis as
\begin{align}
\ket{\uparrow}&=\frac{1}{\sqrt{2}}(\ket{H}+i\ket{V}),\nonumber\\
\ket{\downarrow}&=\frac{1}{\sqrt{2}}(\ket{H}-i\ket{V}).
\end{align}
To detect the polarization of a particle (grin of a cat) in the left or the right arm, a product of the two observables should be introduced, which reads
\begin{align}
\sigma_z^{\mu}=\sigma_z\otimes\Pi_\mu.
\end{align}

According to the Eq.~(\ref{eq:weak}), the weak values of the spatial positions of the particle are measured to be
\begin{align}
\langle\Pi_{\mu}\rangle_w=\delta_{\mu L},
\end{align}
where the $\delta$ symbol is the Kronecker-$\delta$ function, and it means that the particle has traveled through the
left arm.

The weak values of polarization positions of the particle as
\begin{align}
\langle\sigma_z^{\mu}\rangle_w=\delta_{\mu R},
\end{align}
which means that the polarization traveled through the
right arm.

Based on the above result, $\langle\Pi_{L}\rangle_w=\langle\sigma_z^{R}\rangle_w=1$, $\langle\Pi_{R}\rangle_w=\langle\sigma_z^{L}\rangle_w=0$, the null weak value means that we cannot observe the photon in the left arm and cannot observe the polarization property in the right arm. This means the photon is constrained to the right arm and the circular polarization property is constrained to the left arm. The two degrees of freedom location and polarization of a photon are decoupled: the cat and its grin is discovered at different places.

The thought experiment of Quantum Cheshire Cat can be realized by the scheme of Mach-Zehnder interferometer and respectively separated a photon and its circular polarization in the right and left arm.

\section{Two entangled particles can lose their properties}\label{2ecc}

In the previous section, we have seen that a physical
property can be decoupled from a quantum object, i.e., the grin of a cat is separated from the cat itself.
Recently, D. Das {\it et al.} considered two Quantum Cheshire Cats and proposed that two quantum object can not only decouple their physical property, but also swap and re-couple them with the other quantum object~\cite{das2020can}.
This means two Quantum Cheshire Cats can exchange their grins without meeting each other. Soon after, an experimental verification was performed~\cite{liu2020experimental}.

%The polarization property of a photon can be decoupled from the photon itself, i.e., the grin of a cat can be separated from the cat itself. Recently, et al.

In previous work of Quantum Cheshire Cats, the particles are not entangled at the beginning.
This gives rise to a natural question: can two entangled particles lose their properties? The answer is in the positive.
Here, we find that under some appropriate choices of pre- and post-selected states, two initially entangled photons can also decouple from their polarization property.

Here, we also use the composite system of path and polarization as $\ket{\cdot\cdot}_{path} \ket{\cdot\cdot}_{pol}$.
Firstly, we need to choose the pre-selected state as
%\begin{align}\label{eq:2cat-1}
%\ket{\Psi_0}=\frac{1}{\sqrt{2}}(\ket{L_1}\ket{H_1}\ket{R_2}\ket{H_2}+\ket{R_1}\ket{H_1}\ket{L_2}\ket{H_2}),
%\end{align}
\begin{align}\label{eq:2cat-1}
\ket{\Psi_0}=\frac{1}{\sqrt{2}}(\ket{L_1}\ket{R_2}+\ket{R_1}\ket{L_2})\ket{H_1}\ket{H_2},
\end{align}
the post-selected state should be selected as
%\begin{align}\label{eq:2cat-2}
%|\Psi_f\rangle&=\frac{1}{\sqrt{3}}(-i\ket{L_1H_1R_2H_2}+\ket{R_1H_1L_2V_2}+\ket{R_1V_1L_2H_2}).
%\end{align}
\begin{align}\label{eq:2cat-2}
|\Psi_f\rangle&=\frac{1}{\sqrt{3}}(-i\ket{L_1R_2H_1H_2}+\ket{R_1L_2H_1V_2}+\ket{R_1L_2V_1H_2}).
\end{align}

Let us define various operators that measure two photons and their polarizations in which way. For two photons, we also use the projection operators
\begin{align}\label{eq:2cat-3}
\Pi_{\mu_{i}}=\ket{\mu_i}\bra{\mu_i},
\end{align}
to determine whether the photons in the right and left arms. Here, $\mu\in\{L, R\}$ denotes possible right and left paths, $i\in\{1, 2\}$ labels the photons. For example, $\Pi_{L_{1}}$ means whether the first photon is in its left path.

For the polarizations of the two photons, we use the product operators
\begin{align}\label{eq:2cat-4}
\sigma_z^{\mu_i}=\sigma_z^i\otimes\Pi_{\mu_i},
\end{align}
to determine whether the polarization in the right and left arms. Here, the circular polarization observable $\sigma_z^i=\ket{\uparrow}_i\bra{\uparrow}-\ket{\downarrow}_i\bra{\downarrow}$, the basis $\ket{\uparrow}_i=\frac{1}{\sqrt{2}}(\ket{H}_i+i\ket{V}_i)$, $\ket{\downarrow}_i=\frac{1}{\sqrt{2}}(\ket{H}_i-i\ket{V}_i)$.

We find that the weak values of the spatial positions of two photons are measured to be
\begin{align}\label{eq:2cat-5}
\langle\Pi_{\mu_{i}}\rangle_w=\delta_{\mu, L}\delta_{i, 1}+\delta_{\mu, R}\delta_{i, 2},
\end{align}
where the $\delta$ symbol is the Kronecker-$\delta$ function.
It means that the first photon has traveled through his left arm, the second photon passed through his right arm.

The weak values of polarization positions of two particles as
\begin{align}\label{eq:2cat-6}
\langle\sigma_z^{\mu_i}\rangle_w=\delta_{\mu, R}\delta_{i, 1}+\delta_{\mu, L}\delta_{i, 2},
\end{align}
which means that the circular polarization of the first photon traveled through the right arm, the polarization of the second photon traveled through the left arm.

According to the above weak values Eqs.~(\ref{eq:2cat-5}) and (\ref{eq:2cat-6}),
%i.e., $\langle\Pi_{L_{1}}\rangle_w=\langle\sigma_z^{R_2}\rangle_w=1$, $\langle\Pi_{R_{1}}\rangle_w=\langle\sigma_z^{\mu_i}\rangle_w=1$
we intuitively find that two entangled cats and their grins were separated. %In fact, two entangled cats can also exchange their grins even if they are not in the same position.
In the next section, we describe the device that how to achieve the counterintuitive phenomenon through weak value measurements.

\section{Implementation for two entangled particles that decouple their polarization and spatial degree of freedom}\label{setup}

\begin{figure}[!h]
	\centering
	\includegraphics[width=10cm]{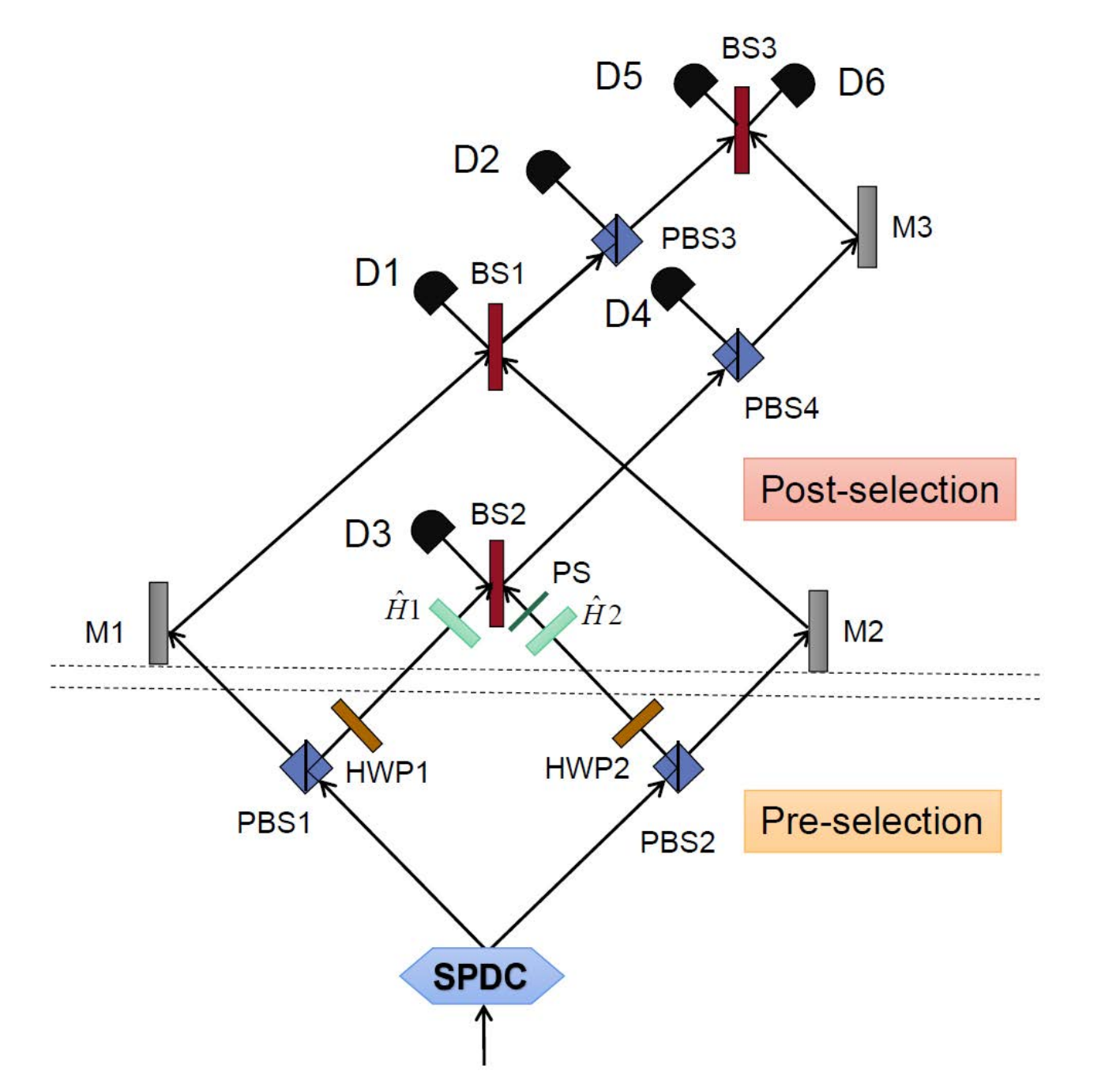}\\
	\caption{Illustration of two entangled particles decouple their polarization.}
	\label{fig:2eqcc}
\end{figure}

Initially, we need to have a pair of entangled Cheshire cats. The entanglement here refers to the spatial position of two cats.
The pre-selection block is constituted by spontaneous parametric down-conversion $SPDC$, two polarization beam-splitter $PBS$, and two half-wave plates $HWP$.

There are two photons from the emission sources, we can label them as $i\in \{1, 2\}$. Now, let them perform the operation of the spontaneous parametric down-conversion $SPDC$, so that we get the polarization entangled photons
\begin{align}
\ket{H_1}\ket{V_2}+\ket{V_1}\ket{H_2},
\end{align}
and let them simultaneously go through polarization beam-splitter $PBS_1$ and $PBS_2$. The $PBS$ always transmits the horizontal polarization state $\ket{H}$ and always reflects the vertical polarization state $\ket{V}$, say. It then transforms  to the state
\begin{align}
%\ket{L_1}\ket{H_1}\ket{R_2}\ket{V_2}+\ket{R_1}\ket{V_1}\ket{L_2}\ket{H_2},
\ket{L_1}\ket{R_2}\ket{H_1}\ket{V_2}+\ket{R_1}\ket{L_2}\ket{V_1}\ket{H_2},
\end{align}
where we define $\ket{L_i}$ and $\ket{R_i}$ as two orthogonal path states, i.e., the left and the right arm of the photon $i$ after passing through $PBS$. Afterwards, the photons pass through the half-wave plates $HWP_1$ and $HWP_2$. The $HWP$ flips the polarization of the photon from $\ket{H}$ to $\ket{V}$ and vice versa.
Then, the preselected state, i.e., two photons entangled state %$\ket{\Psi_0}=\frac{1}{\sqrt{2}}(\ket{L_1}\ket{H_1}\ket{R_2}\ket{H_2}+\ket{R_1}\ket{H_1}\ket{L_2}\ket{H_2})$
$\ket{\Psi_0}=\frac{1}{\sqrt{2}}(\ket{L_1}\ket{R_2}+\ket{R_1}\ket{L_2})\ket{H_1}\ket{H_2}$
was prepared.

%There are two photons from two emission sources, we can label them as $i\in \{1, 2\}$, two of them only have the horizontal polarization state $\ket{H_i}$. Now, let them simultaneously through a 50:50 beam splitter (BS) as the setup~.
%Denoting $\ket{L_i}$ and $\ket{R_i}$ as two orthogonal path states, i.e., the left and the right arm  of the photon $i$ after passing through BS.
%Then, the pre-seledted state, i.e., two photons entangled state %(\ref{eq:2cat-1})
%$\ket{\Psi_0}=\frac{1}{\sqrt{2}}(\ket{L_1}\ket{H_1}\ket{R_2}\ket{H_2}+\ket{R_1}\ket{H_1}\ket{L_2}\ket{H_2})$ were prepared.

Next, we use the measurement strategy---the weak measurement to extract information about this two photons themselves and their polarization.
%Two photons can be detected in the left and the right arms by respectively performing measurements of the observables as Eq.~(\ref{eq:2cat-3}).
To find out which arms the two photons passed through, we perform measurements of the observables as in Eq.~(\ref{eq:2cat-3}). And to find out in which arms the polarizations are detected, we perform measurements of the observables as Eq.~(\ref{eq:2cat-4}).

Finally, we perform post-selection. It is crucial to check whether the state is the post-selected state
%as Eq.~(\ref{eq:2cat-2})
%$|\Psi_f\rangle=\frac{1}{\sqrt{3}}(i\ket{L_1H_1R_2H_2}+\ket{R_1H_1L_2V_2}+\ket{R_1V_1L_2H_2})$,
$|\Psi_f\rangle=\frac{1}{\sqrt{3}}(-i\ket{L_1R_2H_1H_2}+\ket{R_1L_2H_1V_2}+\ket{R_1L_2V_1H_2})$,
which lead to interesting consequences. Such a post-selection block is realized by the optical setup as shown in Fig. \ref{fig:2eqcc}.

The post-selection block is constituted by beam-splitters $BS_i (i=1, 2, 3)$, the Hadamard gates $\hat{H_i} (i=1, 2)$, %half-wave plates $HWP_i (i=...)$,
phase-shifters $PS$, polarization beam-splitters $PBS_i (i=3, 4)$, and detectors $D_i (i=1, 2, 3, 4, 5, 6)$ and mirrors $M_i, (i=1, 2, 3)$.

The right arm of the 1-photon and the left arm of the 2-photon have a $\hat{H}$ operator, respectively. The $\hat{H_i}$ cause the transformations
\begin{align}
\ket{H_i}+\ket{V_i}\rightarrow\ket{H_i},
\end{align}%$\ket{H_i}+\ket{V_i}\rightarrow\ket{H_i}$,
where $i=1, 2$.

If we mapping the basis $\{\ket{H}, \ket{V}\}$ to $\{\ket{0}, \ket{1}\}$, then, the operator $\hat{H}$ is nothing but the Hadamard gate
\begin{align}
\hat{H}=\frac{1}{\sqrt{2}}\left(
  \begin{array}{cc}
    1 & 1 \\
    1 & -1 \\
  \end{array}
\right).
\end{align}
Then, phase-shifters $PS$ add a phase-factor $i$ to the left arm of the 2-photon.

Next, by adjusting the beam-splitters $BS_1$, we choose the state $i\ket{L_1}\ket{R_2}$, and let it go through to polarization beam-splitters $PBS_3$. Meanwhile, the detector $D_1$ will have a click for other state generated on $BS_1$.
As mentioned before, polarization beam-splitters $PBS$ transmit the horizontal polarization state $\ket{H_i}$ and reflect the vertical polarization state $\ket{V_i}$. Then, if the polarization state is $\ket{V_i}$, the detector $D_2$ will have a click. Otherwise, particles with horizontal polarization $\ket{H_i}$ move on $BS_3$.
Likewise, the beam-splitter $BS_2$ allows only the photon in the state $\ket{R_1}\ket{L_2}$ towards the polarization beam-splitters $PBS_4$. And if any other state is incident on $BS_4$, the detector $D_3$ will have a click. The $PBS_4$ is chosen to reflect $\ket{V_i}$, and if so, the detector $D_4$ will have a click. Otherwise, the state $\ket{R_1}\ket{L_2}$ with horizontal polarization $\ket{H_i}$ will move towards the beam-splitters $BS_3$.
Finally, by adjusting the beam-splitter $BS_3$, the state $-i\ket{L_1}\ket{R_2}+2\ket{R_1}\ket{L_2}$ is chosen with
detector $D_5$, while any other state is sent towards the detector $D_6$.

We see that if and only if there is a click of the detector $D_5$, do we obtain a successful post-selection. On the contrary, any clicks in the detectors $D_1, D_2, D_3, D_4, D_6$ signifies that the post-selection was unsuccessful.

\section{The general two entangled Quantum Cheshire Cat scheme}\label{general}

In the above scenario, the pre-selected state $\ket{\Psi_0}$ implicit a maximum entanglement state, its robustness of noisy is relatively weak in experiment. Here we consider the more general case.

More generally, we can choose the general entangled pre-selected state as
\begin{align}\label{eq:g-1}
%\ket{\Psi_0'}=\cos{\theta}\ket{L_1}\ket{H_1}\ket{R_2}\ket{H_2}+\sin{\theta}\ket{R_1}\ket{H_1}\ket{L_2}\ket{H_2},
\ket{\Psi_0'}=(\cos{\theta}\ket{L_1}\ket{R_2}+e^{i\phi}\sin{\theta}\ket{R_1}\ket{L_2})\ket{H_1}\ket{H_2},
\end{align}
and the post-selected state can be selected as
\begin{align}\label{eq:g-2}
%|\Psi_f'\rangle&=\frac{1}{\sqrt{3}}(-i\ket{L_1H_1R_2H_2}+e^{i\phi} \cot{\theta}\ket{R_1H_1L_2V_2}\nonumber\\
%&+e^{i\phi} \cot{\theta}\ket{R_1V_1L_2H_2}).
|\Psi_f'\rangle&=\frac{1}{\sqrt{3}}(-i\ket{L_1R_2H_1H_2}+e^{i\phi} \cot{\theta}\ket{R_1L_2H_1V_2}\nonumber\\
&+e^{i\phi} \cot{\theta}\ket{R_1L_2V_1H_2}).
\end{align}

Various operators are still chosen as Eqs.~(\ref{eq:2cat-3}) and (\ref{eq:2cat-4}). By the calculation, we can obtain the weak values of the spatial positions of two photons
\begin{align}
\langle\Pi_{L_{1}}\rangle_w&=1,~~~\langle\Pi_{R_{1}}\rangle_w=0,\nonumber\\
\langle\Pi_{L_{2}}\rangle_w&=0,~~~\langle\Pi_{R_{2}}\rangle_w=1,
\end{align}
and the weak values of the polarization positions
\begin{align}
\langle\sigma_z^{L_{1}}\rangle_w&=0,~~~\langle\sigma_z^{R_{1}}\rangle_w=1,\nonumber\\
\langle\sigma_z^{L_{2}}\rangle_w&=1,~~~\langle\sigma_z^{R_{2}}\rangle_w=0.
\end{align}
In this way, the polarization of two photons is decoupled from the paths of the photons.

\section{The case of $n$ entangled particles}\label{necc}

Generally, for $n$ entangled particles, it can also be separated from its polarization. Here, we come up with a scenario.

For convenience, we define
\begin{align}
\ket{L}\rightarrow\ket{0},~~~~\ket{R}\rightarrow\ket{1},\\
\ket{H}\rightarrow\ket{0},~~~~\ket{V}\rightarrow\ket{1}.
\end{align}

The entangled pre-selected state is  chosen as
\begin{align}\label{eq:necc-1}
\ket{\Psi_0^n}=\frac{1}{\sqrt{2}}(\ket{[\sum_{k=1}^{\lfloor{\frac{n}{2}}}2^{2n-2k}]_b}+\ket{[\sum_{j=1}^{\lfloor{\frac{n+1}{2}}}2^{2n-2j+1}]_b}),
\end{align}
and the post-selected state is selected as
\begin{align}\label{eq:necc-2}
|\Psi_f^n\rangle&=\frac{1}{\sqrt{n+1}}(-i\ket{[\sum_{k=1}^{\lfloor{\frac{n}{2}}}2^{2n-2k}]_b}+\ket{[\sum_{j=1}^{\lfloor{\frac{n+1}{2}}}2^{2n-2j+1}+1]_b}\nonumber\\
&+\sum_{l=0}^{n-2}\ket{[\sum_{j=1}^{\lfloor{\frac{n+1}{2}}}2^{2n-2j+1}+2^l+1]_b}),
\end{align}
where $\lfloor{x}$ is the floor function, which map $x$ to the greatest integer less than or equal to $x$; and $[y]_b$ represents the binary number of $y$.

As before, in order to measure the positions of the particles and their polarizations, we choose the operator as Eqs.~(\ref{eq:2cat-3}) and (\ref{eq:2cat-4}). We can then obtain weak values,
\begin{align}\label{eq:2cat-5}
\langle\Pi_{\mu_{i}}\rangle_w&=\delta_{\mu, L}\delta_{i, 2m-1}+\delta_{\mu R}\delta_{i, 2m},\\
\langle\sigma_z^{\mu_i}\rangle_w&=\delta_{\mu, R}\delta_{i, 2m-1}+\delta_{\mu L}\delta_{i, 2m},
\end{align}
i.e., $2m-1$-th photon pass through its left arm, and its polarization pass through its right arm; while $2m$-th photon pass through its right arm, whose polarization pass through its left arm; $m=1, 2, \cdots$

\section{Conclusion and Discussion}

In summary, we have proposed a thought experiment in which the circular polarizations of two entangled photons can be decoupled from their positions. %Two entangled photons decoupled with their polarization, respectively, and then, they can capture the polarization of the other photon.
It is significant that the separation of the polarization of two entangled photons. Especially in the previous, people have done many works on the nature of entanglement, but they didn't take into account the separation of attributes.
In this work, we mainly used the advantages of weak value, through designing a pre-and post-measurement setup, we can achieve the counterintuitive phenomenon. More general, two arbitrarily entangled Quantum Cheshire Cat scenario and $n$ entangled Quantum Cheshire Cat scenarios were given.
Our results will deepen the understanding of the basic concepts of quantum measurement, and also gradually advance the theoretical framework of the fundamental issues of quantum mechanics. In the future, it is expected to be realized experimentally.

\begin{acknowledgments}
J.Z. was supported by the China Scholarship Council (No.202006200142). J.L.C. is supported by National Natural Science Foundation of China (Grant Nos. 11875167, 12075001), the Fundamental Research Funds for the Central Universities (Grant No. 3072022TS2503).  K.L.C acknowledges support from the Ministry of Education, Singapore and the National Research Foundation, Singapore.
\end{acknowledgments}

\appendix

\section{The calculation of two entangled Quantum Cheshire Cats}

Assuming that two particles are entangled at the initial moment as Eq.~(\ref{eq:g-1}), how to find the post-selected state, Eq.~(\ref{eq:g-2})? Here we give a simple calculation.

We can suppose the post-selected state (non-normalized) is arbitrary 4-qubit state
\begin{align}
|\tilde{\Psi}_f^2\rangle&=(m(1)^\dagger, m(2)^\dagger, m(3)^\dagger,m(4)^\dagger,m(5)^\dagger,m(6)^\dagger,m(7)^\dagger,\nonumber\\
&m(8)^\dagger,m(9)^\dagger,m(10)^\dagger,m(11)^\dagger,m(12)^\dagger,m(13)^\dagger,m(14)^\dagger,\nonumber\\
&m(15)^\dagger,m(16)^\dagger)^T,
\end{align}
then, the weak values
\begin{align}
(\Pi_{L_1})^w&=\frac{m(5) \cos (x)}{m(9) \sin (x)+m(5) \cos (x)},\nonumber\\
(\Pi_{R_1})^w&=\frac{m(9) \sin (x)}{m(9) \sin (x)+m(5) \cos (x)},\nonumber\\
(\Pi_{L_2})^w&=\frac{m(9) \sin (x)}{m(9) \sin (x)+m(5) \cos (x)},\nonumber\\
(\Pi_{R_2})^w&=\frac{m(5) \cos (x)}{m(9) \sin (x)+m(5) \cos (x)},\nonumber\\
(\sigma_z^{L_1})^w&=\frac{i m(7) \cos (x)}{m(9) \sin (x)+m(5) \cos (x)},\nonumber\\
(\sigma_z^{R_1})^w&=\frac{i m(11) \sin (x)}{m(9) \sin (x)+m(5) \cos (x)},\nonumber\\
(\sigma_{z}^{L_2})^w&=\frac{i m(10) \sin (x)}{m(9) \sin (x)+m(5) \cos (x)},\nonumber\\
(\sigma_{z}^{R_2})^w&=\frac{i m(6) \cos (x)}{m(9) \sin (x)+m(5) \cos (x)}.
\end{align}

According to the Quantum Cheshire Cat thought experiment, they must be satisfied $\langle\Pi_{L_{1}}\rangle_w=1, \langle\Pi_{R_{1}}\rangle_w=0,
\langle\Pi_{L_{2}}\rangle_w=0, \langle\Pi_{R_{2}}\rangle_w=1, \langle\sigma_z^{L_{1}}\rangle_w=0, \langle\sigma_z^{R_{1}}\rangle_w=1,
\langle\sigma_z^{L_{2}}\rangle_w=1, \langle\sigma_z^{R_{2}}\rangle_w=0$, then, we have
\begin{align}
m(5)=i,~~~m(10)=m(11)=e^{i\phi}\cot{\theta},
\end{align}
and other items are 0.

Then, we get the post-selected state as Eq.~(\ref{eq:g-2}). Further more, when $\phi=0, \theta=\pi/4$, the pre- and post-selected state as Eqs.~(\ref{eq:2cat-1}) and (\ref{eq:2cat-2}).

\section{The calculation of $n$ entangled Quantum Cheshire Cats}

For simplicity, we use $\ket{L}\rightarrow\ket{0}, \ket{R}\rightarrow\ket{1}, \ket{H}\rightarrow\ket{0}, \ket{V}\rightarrow\ket{1}$. Then, the pre- and post-selected states of two entangled Quantum Cheshire Cat in Eqs.~(\ref{eq:2cat-1}) and (\ref{eq:2cat-2}) can be rewritten as
\begin{align}
\ket{\Psi_0^2}=\frac{1}{\sqrt{2}}(\ket{0100}+\ket{1000}),
\end{align}
\begin{align}
|\Psi_f^2\rangle&=\frac{1}{\sqrt{3}}(-i\ket{0100}+\ket{1001}+\ket{1010}).
\end{align}

Notation: in each item of above states, the first half represents the path part, and the second half represents the polarization part, $\ket{\cdot\cdot}_{path} \ket{\cdot\cdot}_{pol}$. For example, $\ket{\Psi_0^2}=\frac{1}{\sqrt{2}}(\ket{0100}+\ket{1000})$ means $\frac{1}{\sqrt{2}}(\ket{\underbrace{01}_{path}\underbrace{00}_{pol}}+\ket{\underbrace{10}_{path}\underbrace{00}_{pol}})$.

Similarly, we can get the pre- and post-selected states of $n=3, 4, 5$ entangled Quantum Cheshire Cat
\begin{align}
\ket{\Psi_0^3}&=\frac{1}{\sqrt{2}}(\ket{010000}+\ket{101000}),\\
|\Psi_f^3\rangle&=\frac{1}{\sqrt{4}}(-i\ket{010000}+\ket{101001}+\ket{101010}+\ket{101100}),
\end{align}

\begin{align}
\ket{\Psi_0^4}&=\frac{1}{\sqrt{2}}(\ket{01010000}+\ket{10100000}),\\
|\Psi_f^4\rangle&=\frac{1}{\sqrt{5}}(-i\ket{01010000}+\ket{10100001}\nonumber\\
&+\ket{10100010}+\ket{10100100}+\ket{10101000}),
\end{align}

\begin{align}
\ket{\Psi_0^5}&=\frac{1}{\sqrt{2}}(\ket{0101000000}+\ket{1010100000}),\\
|\Psi_f^5\rangle&=\frac{1}{\sqrt{6}}(-i\ket{0101000000}+\ket{1010100001}+\ket{1010100010}\nonumber\\
&+\ket{1010100100}+\ket{1010101000}+\ket{1010110000}).
\end{align}

Let's summarize the general formula. For the convenience of finding a general formula, we take each term as binary form, and then replace them with the decimal form. We list various items of the $n$ entangled pre- and post-selected states in Tables.~\ref{tab:pre} and \ref{tab:post}. Intuitively, we can obtain the general formula as Eqs.~(\ref{eq:necc-1}) and (\ref{eq:necc-2}).

\begin{widetext}

\begin{table}[htbp]
    \centering
    \caption{Two items of the $n$ entangled pre-selected state. Where the first row represents the coefficients, the last row are the corresponding decimal form. Whereas, $[x]_b$ means the binary number of $x$.}\label{tab:pre}
    \begin{tabular}{c|c|c}
    \hline\hline
        coe & $\frac{1}{\sqrt{2}}$ & $\frac{1}{\sqrt{2}}$  \\
    \hline
        n=2 &$\ket{01 00}$, $2^2$ & $\ket{10 00}$, $2^3$ \\
    \hline
        n=3 & $\ket{010 000}$, $2^4$ & $\ket{101 000}$, $2^5+2^3$ \\
    \hline
        n=4 & $\ket{0101 0000}$, $2^6+2^4$ & $\ket{1010 0000}$, $2^7+2^5$\\
    \hline
        n=5 & $\ket{01010 00000}$, $2^8+2^6$ & $\ket{10101 00000}$, $2^9+2^7+2^5$\\
    \hline
%        n=6 & $\ket{010101 000000}$, $2^{10}+2^8+2^6$ & $\ket{101010 000000}$, $2^{11}+2^9+2^7$\\
%    \hline
%        n=7 & $\ket{0101010 0000000}$, $2^{12}+2^{10}+2^8$ & $\ket{1010101 0000000}$, $2^{13}+2^{11}+2^9+2^7$\\
%    \hline
%        n=8 & $\ket{01010101 00000000}$, $2^{14}+2^{12}+2^{10}+2^8$ & $\ket{10101010 00000000}$, $2^{15}+2^{13}+2^{11}+2^9$\\
%    \hline
        n & $\ket{0101\cdots 0 000\cdots00}$ & $\ket{1010\cdots 1 000\cdots00}$ \\
    \hline
        n & $\ket{[\sum_{k=1}^{\lfloor{\frac{n}{2}}}2^{2n-2k}]_b}$ &
        $\ket{[\sum_{j=1}^{\lfloor{\frac{n+1}{2}}}2^{2n-2j+1}]_b}$ \\
    \hline\hline
    \end{tabular}
\end{table}

\begin{table}[htbp]
    \centering
    \caption{Various items of the $n$ entangled post-selected state. The first row represents the coefficients and the last row are the corresponding decimal form. Note that $[x]_b$ means the binary number of $x$.}\label{tab:post}
    \begin{tabular}{c|c|c|c|c|c|c|}
    \hline\hline
        coe & $-\frac{i}{\sqrt{n+1}}$ & \multicolumn{5}{|c|}{$\frac{1}{\sqrt{n+1}}$} \\
    \hline
        n=2 & $\ket{01 00}$, $2^2$ & $\ket{10 01}$, $2^3+2^0$ & $\ket{10 10}$ \\
    \hline
        n=3 & $\ket{010 000}$, $2^4$ & $\ket{101 001}$, $2^5+2^3+2^0$ & $\ket{101 010}$ & $\ket{101 100}$ \\
    \hline
        n=4 & $\ket{0101 0000}$, $2^6+2^4$ & $\ket{1010 0001}$, $2^7+2^5+2^0$ & $\ket{1010 0010}$ & $\ket{1010 0100}$ & $\ket{1010 1000}$ \\
    \hline
        n=5 & $\ket{01010 00000}$, $2^8+2^6$ & $\ket{10101 00001}$, $2^9+2^7+2^5+2^0$ & $\ket{10101 00010}$ & $\ket{10101 00100}$ & $\ket{10101 01000}$ & $\ket{10101 10000}$ \\
    \hline
        n & $\ket{[\sum_{k=1}^{[\frac{n}{2}]_2}2^{2n-2k}]_b}$ & $\ket{[\sum_{j=1}^{[\frac{n+1}{2}]}2^{2n-2j+1}+1]_b}$ & \multicolumn{4}{|c|}{$\ket{[\sum_{j=1}^{[\frac{n+1}{2}]}2^{2n-2j+1}+1+2^l]_b}, l=0,1,\cdots,n-2$} \\
    \hline\hline
    \end{tabular}
\end{table}

\end{widetext}

\providecommand{\noopsort}[1]{}\providecommand{\singleletter}[1]{#1}%
%

%\bibliography{QCC}
%\bibliographystyle{apsrev4-1}

\end{document}